\documentclass{iau} 
\usepackage{graphicx}

\title[Testing SiO Maser Polarization Theories] 
{Constraining Theories of SiO Maser Polarization:\\ Analysis of a $\pi/2$ EVPA Change}

\author[T. L. Tobin, A. J. Kemball, \& M. D. Gray]   
{T. L. Tobin$^{1,2}$, A. J. Kemball$^1$
 \and M. D. Gray$^3$}

\affiliation{$^1$Department of Astronomy, University of Illinois at Urbana-Champaign\\1002 W. Green Street, Champaign, IL 61801, USA
\\ $^2$email: {\tt tltobin2@illinois.edu} \\[\affilskip]
$^3$Jodrell Bank Centre for Astrophysics, Alan Turing Building, University of Manchester\\Manchester M13 9PL, UK}

\pubyear{2017}
\volume{336}  
\jname{Astrophysical Masers: Unlocking the Mysteries of the Universe}
\editors{A. Tarchi, M.J. Reid \& P. Castangia, eds.}
\begin{document}

\maketitle

\begin{abstract}
The full theory of polarized SiO maser emission from the near-circumstellar 
environment of Asymptotic Giant Branch stars has been the subject of debate, 
with theories ranging from classical Zeeman origins to 
predominantly non-Zeeman anisotropic excitation or propagation effects. 
Features with an internal electric vector position angle 
(EVPA) rotation of $\sim \pi /2$ offer unique constraints on theoretical 
models. In this work, results are presented for one such feature that 
persisted across five epochs of SiO $\nu=1, J=1-0$ VLBA observations of TX 
Cam. We examine the fit to the predicted dependence of linear polarization and 
EVPA on angle ($\theta$) between the line of sight and the magnetic field 
against theoretical models. We also present results on the dependence of 
$m_c$ on $\theta$ and their theoretical implications.  Finally, we discuss 
potential causes of the observed differences, and continuing work.
\end{abstract}

\firstsection 
\section{Introduction}

Although theories have endeavored to explain the polarization of SiO masers originating from the near circumstellar environments (NCSE) of Asymptotic Giant Branch (AGB) stars, no theoretical consensus has yet been reached. Prominent theories as to the origin of SiO $\nu=1, J=1-0$ maser polarization ascribe it to the local magnetic field (\cite[Goldreich et al. 1973]{GKK} , \cite[Elitzur 1996]{eli96}) or a change in the anisotropy of pumping radiation conditions or other non-Zeeman effects (\cite[Asensio Ramos et al. 2005]{aram05}, \cite[Watson 2009]{watson09}).

\begin{figure}[t]
\begin{center}
 \includegraphics[width=3.4in]{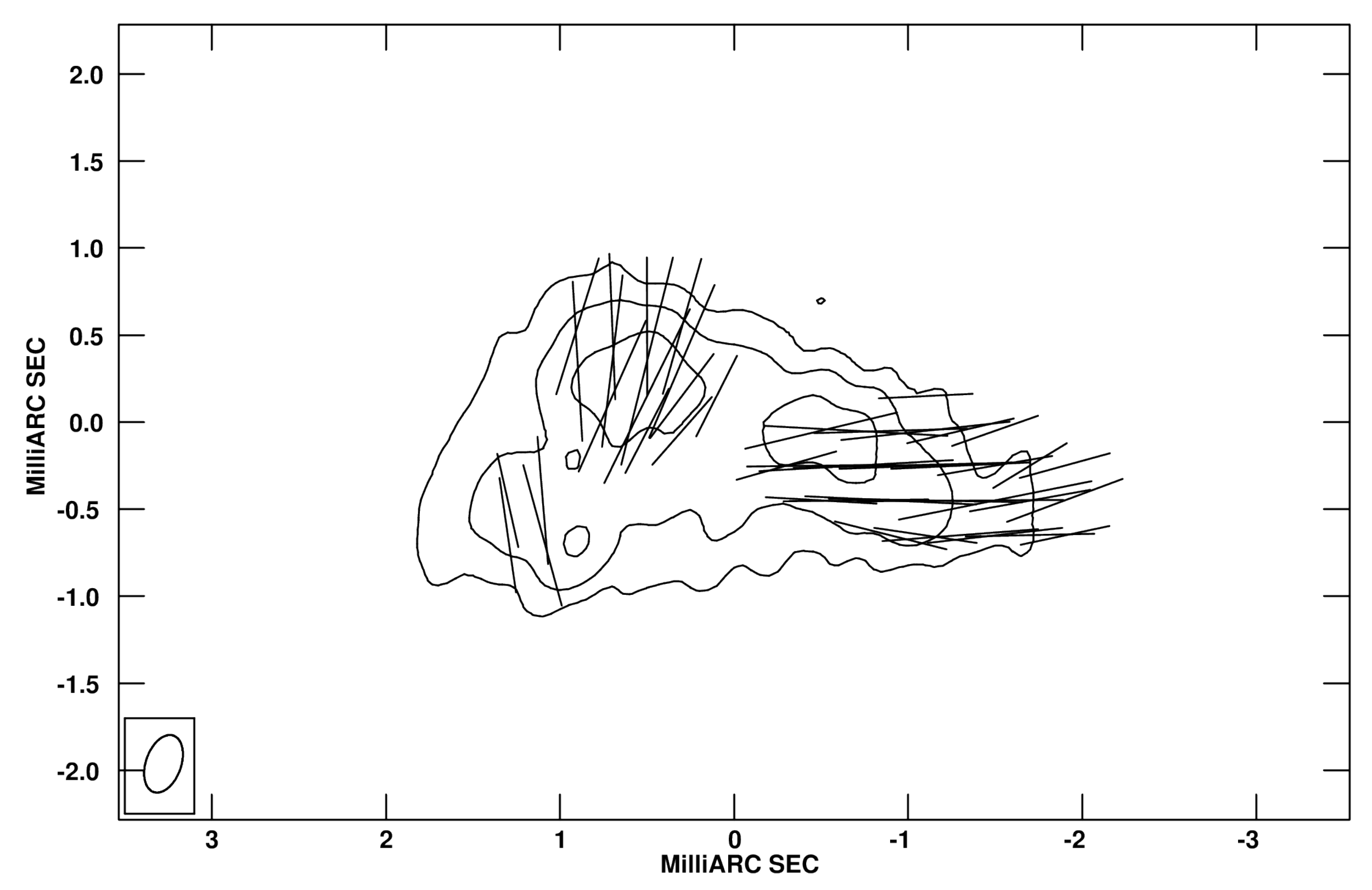} 
 \caption{Target maser feature in epoch BD46AO. Contours denote frequency-averaged Stokes I with levels of $\{-10,-5,5,10,20,40,80,160,320\} \times \sigma$, where $\sigma_{AO}=1.6430$ mJy beam$^{-1}$. Vectors denote the frequency-averaged linear polarization, with 1 mas in vector length corresponding to 4 mJy beam$^{-1}$.\label{ao_feat}}
   \label{fig1}
\end{center}
\end{figure}

However, these theories differ in their ability to explain rotations of the EVPA by $\sim \pi/2$ within a single maser feature. In some theories, such as \cite[Goldreich et al. (1973)]{GKK} (hereafter GKK), the Electric Vector Position Angle (EVPA) is governed by the angle, $\theta$, between the magnetic field and the line of sight. When $\theta$ is small, the linear polarization would be parallel to the projected magnetic field. However, when $\theta$ becomes larger than the Van Vleck angle ($\sim 55 ^\circ$), the polarization would be perpendicular to the projected magnetic field. In this case, a rotation of the EVPA across a feature could be due to a slight change in the direction of the magnetic field with respect to the line of sight, spanning the Van Vleck angle. 

%
%

\begin{figure}[t!]
\begin{center}
 \includegraphics[width=5.4in]{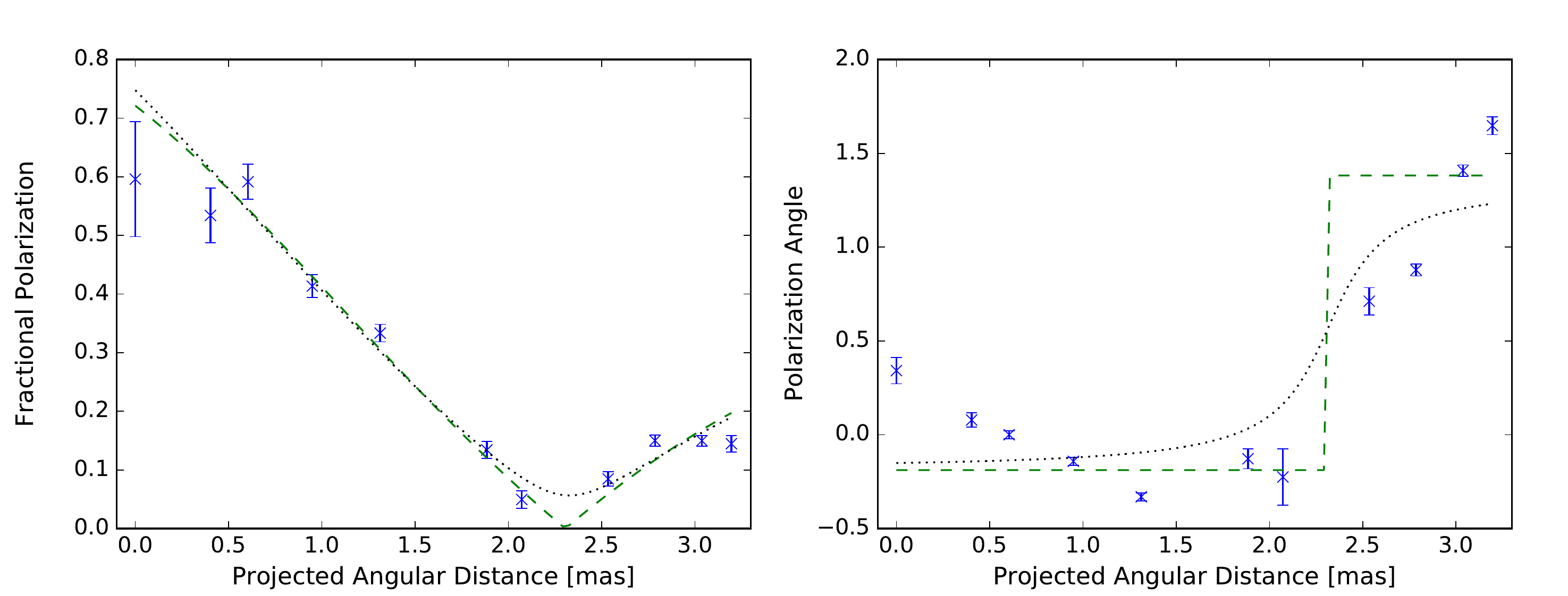} 
 \caption{Fractional linear polarization (left) and relative EVPA (right) as a function of projected angular distance for epoch BD46AP. In both plots, 'X's with errors indicate the data. (Note: errors from absolute calibration of EVPA are not included, since the shape of the profile is the focus.) The best fit of the fractional linear polarization from GKK with $K=0$ is the dashed line, while the fit with non-zero K is the dotted line.}
   \label{fig2}
\end{center}
\end{figure}

This rotation could also be due to a change in the direction of the projected magnetic field across the spatial extent of the maser feature in the image plane. In this case, the EVPA would again be defined by the direction of the projected magnetic field in the sky, but the angle of the projected magnetic field would rotate within the masing material (\cite[Soker \& Clayton 1999]{sokerclayton}). 

Alternately, if the polarization is mainly governed by anisotropy of pumping radiation conditions, such a rotation could indicate a change in those conditions across the maser feature (\cite[Asensio Ramos et al. 2005]{aram05}). Here, we discuss our analysis of a maser feature with an internal EVPA rotation of $\sim \pi/2$ that persists across five epochs of observations, and our application of several tests of SiO maser polarization theories to the feature.

\section{Observations\label{obs}}
For this analysis, we used five epochs of the long-term, full-polarization SiO 
$\nu=1 J=1-0$ (43 GHz) VLBA campaign of the Mira variable, TX Cam. These 
observations have been previously analyzed for total intensity and kinematics 
(\cite[Diamond \& Kemball 2003]{diakem03}, \cite[Gonidakis et al. 2010]
{GDK10}), and linear polarization (\cite[Kemball et al. 2009]
{kemb09}). This work focuses on epoch codes BD46AN, BD46AO, BD46AP, BD46AQ, 
and BD46AR. For further information on the observations themselves, please see 
\cite[Diamond \& Kemball (2003)]{diakem03}. 

In addition, the linear polarization of our target maser feature was analyzed 
for one epoch (BD46AQ) in \cite[Kemball et al. (2011)]{kdrgx11}. As was done 
in that work, we reduce the data using the method described in 
\cite[Kemball \& Richter (2011)]{kembrich11}, to obtain accurate measurements 
of the low levels of circular polarization. This work 
in particular expands on previous work by increasing the number of 
epochs analyzed with accurate circular polarization and applying additional tests of maser polarization theory to the data. 

\section{Discussion}
{\underline{\it GKK and Linear Polarization}}. \cite[GKK]{GKK}  cite an asymptotic solution for fractional Q and U polarization, Y and Z, respectively, in the regime $\Delta \omega \gg g \Omega \gg R \gg \Gamma$, as 
$Y=\frac{3 \sin^2 \theta - 2}{3 \sin^2 \theta},  Z=K$ for $\sin^2\theta \geq \frac{1}{3}$, and $Y=-1, Z=0$ for $\sin^2 \theta \leq \frac{1}{3}$,
where Stokes V is assumed to be zero and K is some number such that $Y^2 + Z^2 
\leq 1$. Typical applications of this theory assume $K=0$. In the first plot in Figure \ref{fig2}, we fit the 
predicted linear polarization fraction, $m_l$, as a function of projected 
angular distance to the prediction by \cite[GKK]{GKK}. To do this, we assumed 
$\theta$ was a quadratic function of projected angular distance, $d$, and fit 
for the first- and second-order coefficients, and $d_f$, the value of $d$ at which $\theta$ is the Van Vleck angle, following \cite[Kemball et al. (2011)]{kdrgx11}: $\theta = p_0 ( d^2 - d_f^2 ) + p_1 (d - d_f) + \arcsin \sqrt{2/3}$. For completeness, we fit for both $K=0$ and non-zero $K$. Notably, while this profile fit some epochs better than others, it provides a 
remarkably good fit.

The expected EVPA profiles are derived directly from the best fit functions to $m_l$, and then fit to the measured EVPA with a simple vertical offset,
as we are not accounting for absolute EVPA. The results can be seen in the second plot
in Figure \ref{fig2}.

In this case, $K=0$ \cite[GKK]{GKK} predicts a that the EVPA profile will be a 
strict step function, whereas the data show a smooth rotation of the linear 
polarization. 
Generally, adding a non-zero K smooths out the rotation. However, it also 
causes the extremal angles to be approached asymptotically and can result in 
less of a net rotation. In contrast, most epochs of our data actually show a 
rotation of slightly more than $\pi/2$. Notably, such an investigation of EVPA rotation was also conducted by \cite[Vlemmings \& Diamond (2006)]{vlemdiam} for an H$_2$O maser of W43A, although the complex Zeeman structure of the water transition prevents the results from being directly analogous.


\begin{figure}[t!]
\begin{center}
 \includegraphics[width=5.4in]{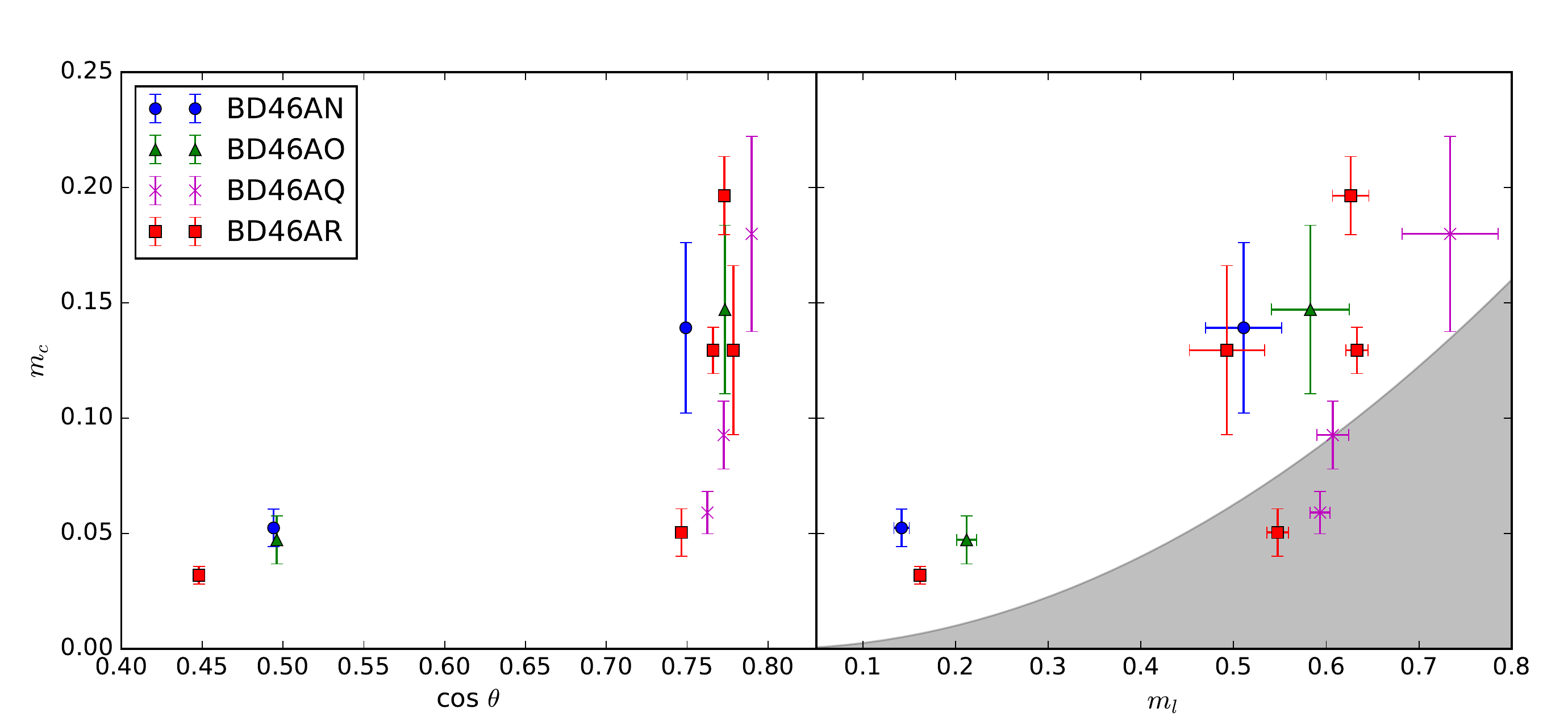} 
 \caption{Fractional circular polarization, $m_c$, as a function of $\cos \theta$, as determined by the K=0 GKK fit (left) and fractional linear polarization, $m_l$ (right). Points shown have $m_c$ $S/N > 3$. Grey shading in right plot denotes region consistent with $m_c < m_l^2/4$.\label{fig3}}
\end{center}
\end{figure}

{\underline{\it {Zeeman Circular Polarization}}. Although \cite[GKK]{GKK} 
assumed Stokes $V=0$, others have expanded on this theory by deriving the 
behavior of circular polarization due to Zeeman splitting. 
\cite[Elitzur (1996)]{eli96} predicted that the $m_c \propto 1 / \cos \theta$. 
\cite[Gray (2012)]{gray12} predicted that $m_c$ is roughly proportional to 
$\cos \theta$ but it may not be a purely linear relation. Finally, 
\cite[Watson \& Wyld (2001)]{watwyld01} predicted a more complex, peaked 
function for $m_c(\cos \theta)$. The left plot in Figure \ref{fig3} shows measured $m_c$ as a function of $\cos 
\theta$ as determined from the $K=0$ \cite[GKK]{GKK} fit to the linear 
polarization fraction profile. Although there is scatter at higher $\cos 
\theta$, our data appears most consistent with the prediction from \cite[Gray 
(2012)]{gray12}.

{\underline{\it {Non-Zeeman Circular Polarization}}. \cite[Wiebe \& Watson 
(1998)]{wiewat98} suggested that circular polarization may be a result of 
conversion from linear polarization due to non-Zeeman effects such as changing optical 
axes in the medium or a change in the magnetic field orientation along the 
line of sight. This type of non-Zeeman circular polarization would be limited 
by $m_c < m_l^2 / 4$ (\cite[Wiebe \& Watson 1998]{wiewat98}). As shown in the second plot of Figure \ref{fig3}, the vast majority of our 
data are not consistent with this limit. \cite[Wiebe \& Watson (1998)]
{wiewat98} suggest that, individual points may fall outside this limit, but 
the average values should be consistent if the circular polarization is 
arising via this mechanism. Even averaging our values over epoch, not a single 
epoch is consistent with this limit. This is consistent with the findings of 
\cite[Cotton et al. (2011)]{cotton11}.

{\underline{\it Alternative Theories}}. Other explanations of this 
EVPA rotation include a curvature of the magnetic field itself within the  masing material. Local changes in the direction of the 
magnetic field such as this have been predicted in \cite[Soker \& Clayton 
(1999)]{sokerclayton}. In this case, the EVPA would be tracing the projected 
magnetic field as it rotates. However, if this was the case, we wouldn't 
expect to see the $m_l$ profile that so closely resembles \cite[GKK]{GKK}.

Another possibility is that, instead of resulting from interaction with 
magnetic fields, the change in EVPA is a result of changing anisotropy 
conditions. \cite[Asensio Ramos et al. (2005)]{aram05} propose that a change 
the anisotropic pumping conditions could cause a rotation in the EVPA. 
However, a more extensive parameter space and lack of concurrent $m_l$ 
predictions prevent application of a definitive test.



\acknowledgments
This material is based upon work supported by the National Science Foundation Graduate Research Fellowship Program under Grant No. DGE - 1144245.

\end{document}